\documentclass{aa}

\usepackage{graphicx}
\usepackage{txfonts}
\usepackage{natbib}
\bibpunct{(}{)}{;}{a}{}{,}

\usepackage[colorlinks=true, citecolor=blue]{hyperref}

\begin{document}

   \title{Investigating the interstellar dust through the Fe K-edge}
   
   \titlerunning{The Fe K-edge}
   \authorrunning{D. Rogantini, E. Costantini et al.}

   \subtitle{}

   \author{D. Rogantini,
          \inst{1,2}
          E. Costantini,\inst{1,2}
          S.T. Zeegers,\inst{1,3}
          C.P. de Vries,\inst{1}
          W. Bras,\inst{4}
           F. de Groot,\inst{5}
           H. Mutschke,\inst{6}
           \and
          L.B.F.M. Waters\inst{1,2}
          }

   \institute{SRON Netherlands Institute for Space Research, Sorbonnelaan 2, 3584 CA Utrecht, the Netherlands\\
              \email{d.rogantini@sron.nl}
             \and 
             Anton Pannekoek Astronomical Institute, University of Amsterdam, P.O. Box 94249, 1090 GE Amsterdam, the Netherlands
             \and
             Leiden Observatory, Leiden University, PO Box 9513, NL-2300 RA Leiden, the Netherlands
             \and
             Netherlands Organisation for Scientific Research (NWO) DUBBLE@ESRF, CS 4022, 38043, Grenoble Cedex 9, France
             \and
             Debye Institute for Nanomaterials Science, Utrecht University, Universiteitweg 99, 3584 CG Utrecht, the Netherlands
             \and
             Astrophysikalisches Institut und Universit$\ddot{\text{a}}$ts-Sternwarte (AIU), Schillerg{\"a}{\ss}chen 2-3, 07745 Jena, Germany
             }
    \date{}

 
  \abstract
   {The chemical and physical properties of interstellar dust in the densest regions of the Galaxy are still not well understood. X-rays provide a powerful probe since they can penetrate gas and dust over a wide range of column densities (up to $10^{24}\ \rm{cm}^{-2}$). The interaction (scattering and absorption) with the medium imprints spectral signatures that reflect the individual atoms which constitute the gas, molecule, or solid.}
   {In this work we investigate the ability of high resolution X-ray spectroscopy to probe the properties of cosmic grains containing iron. Although iron is heavily depleted into interstellar dust, the nature of the Fe-bearing grains is still largely uncertain.}
   {In our analysis we use iron K-edge synchrotron data of minerals likely present in the ISM dust taken at the European Synchrotron Radiation Facility. We explore the prospects of determining the chemical composition and the size of astrophysical dust in the Galactic centre and in molecular clouds with future X-ray missions. The energy resolution and the effective area of the present X-ray telescopes are not sufficient to detect and study the Fe K-edge, even for bright X-ray sources.}
   {From the analysis of the extinction cross sections of our dust models implemented in the spectral fitting program SPEX, the Fe K-edge is promising for investigating both the chemistry and the size distribution of the interstellar dust. We find that the chemical composition regulates the X-ray absorption fine structures in the post edge region, whereas the scattering feature in the pre-edge is sensitive to the mean grain size. Finally, we note that the Fe K-edge is insensitive to other dust properties, such as the porosity and the geometry of the dust. }
   {}


   \maketitle
%

\section{Introduction}

Iron plays a crucial role in biology and in human life. It is an abundant chemical element on Earth and in the Galaxy. The solar photospheric abundance\footnote{Abundance is given in logarithmic scale $\log A_{\text{Fe}} = \log N_{\text{Fe}}/N_{\text{H}} + 12.0$, where $N_{\text{Fe}}$ is the column density of iron. By definition the solar abundance of hydrogen is exactly $\log A_{\text{H}}$ = 12.0 or $10^6\ $ppm.} of iron is $\log A_{\text{Fe}} = 7.50 \pm 0.05$ or $32\ $ppm \citep{Lodders10} and is consistent with the value derived from meteorites \citep{Anders89}. Iron is by far one of the most dominant species in the solar visible spectrum. This is also true for a large number of stellar spectra. When comparing solar abundances with those measured in other stars, the iron abundance shows a spatial trend along the Galactic plane: it decreases with increasing galactocentric distance ($R_{\text{G}}$). \cite{Genovali14} found a logarithmic iron abundance\footnote{The logarithmic abundance of Fe relative to its solar abundance is defined as $[\text{Fe}/\text{H}]=\log {(N_{\text{Fe}}/N_{\text{H}})} - \log {(N_{\text{Fe}}/N_{\text{H}})}_{\odot}\ .$} gradient of $[\text{Fe}/\text{H}]= 0.57 \pm 0.02 - 0.060 \pm 0.002\ R_{\text{G}}/\text{kpc}$, studying the optical spectra of a large number of Galactic Cepheids.\\
 The abundance of iron in diffuse or molecular clouds determined from spectral lines shows a dramatic shortfall compared with the standard reference value. This apparent shortage, or depletion, of iron and metals (in general) in the gas-phase is interpreted as evidence for their inclusion in interstellar dust grains. In particular, Fe is heavily depleted and the value changes just slightly with the density of the environment. Indeed iron depletion remains high: 89 and 99\% depletions respectively in the warm and the cool phases of the interstellar medium \citep[ISM,][]{Savage96, Jenkins09}.\\ 
Iron is known to be primarily produced in Type I Supernovae (SN Ia) or at the endpoint of the evolution of massive stars that end their life as core collapse supernovae (SNCC). Asymptotic giant branch (AGB) stars are another important source of dust with metal inclusion. Only SNCC and AGB stars are observed to be sources of interstellar dust, since searches for dust in SN Ia have shown that these events are not major providers of fresh dust \citep{Gomez12}. \cite{Dwek16} infers that more than 65\% of the total iron is injected into the ISM in gaseous form by SN Ia. Thus, to explain its depletion, most iron growth needs to occur outside the sources of stellar condensation. Iron must have accreted from the ISM gas by cold accretion onto pre-existing silicate, carbon, or other composite grains \citep{Draine09b}.\\
Although iron is predominantly included in solid grains, the composition of the Fe-bearing grains is largely uncertain. The elements Mg and Si are also highly depleted, and together with Fe, they are the main constituents of cosmic silicates \citep{Henning10}. In the Galactic plane there is growing evidence that silicates are Mg-rich, rather than Fe-rich \citep{Costantini05,Costantini12,Min07,Altobelli16}. \cite{Poteet15} investigated the composition of interstellar dust along the line of sight toward $\zeta$ Ophiuchi. They infer that $\lesssim40\%$ of the available elemental Fe abundance is locked up in silicate grains. Iron is expected to be present in other solid-state species. It could exist in pure metallic nanoparticles or Fe oxides \citep[e.g.][and reference therein]{Kemper02,Lee09,Draine13,Poteet15} or as metallic inclusions in glass with embedded metal and sulphides \citep[GEMS, e.g.][]{Bradley94,Xiang11,Keller13} of interstellar origin.\\
  Iron is highly elusive at longer wavelengths, even at ultraviolet (UV) and infrared (IR) frequencies which are broadly used to investigate interstellar dust. Iron does not show any vibrational modes in the mid-infrared wavelength regions and contributes only to the continuum opacity of dust. However, Fe-bearing oxides exhibit single or multiple vibrational resonances within the $17-28\ \mu$m spectral region, depending on the precise composition of these species. Because their spectral features are blended with the strong blending mode of cosmic silicates, Fe-bearing oxides are not detected in the diffuse ISM, and thus are not usually considered to be a significant component of interstellar dust \citep[e.g.][]{Chiar06}.\\
 
 \noindent 
In addition to the study of the gas phase, X-ray high resolution spectroscopy allows us to study the presence and the abundances of the depleted elements along the line of sight. The following photoelectric K-edges of cosmic elements are located in the X-ray energy range $(0.2-10\  \text{keV})$: C, N, O, Mg, Al, Si, S, Ca, Fe. In particular, iron shows two different absorption edges: the Fe K-edge (when a photon has the same energy of the K shell electron of an atom) at $7.112\ $keV \citep{Bearden67} and three Fe L-edges (excitation of a 2s or 2p electron) located in the soft X-rays at $ 0.846,\ 0.721,\ \text{and}\ 0.708\ $keV \citep[L$_{\text{I}}$, L$_{\text{II}}$, and L$_{\text{III}}$, respectively,][]{Bradley94}.\\
  In the post-edge region are located characteristic features which can be up to few electronvolt in width and are known as X-ray absorption fine structures (XAFS, see \cite{Bunker10} for a theoretical explanation). When a photon excites a core electron ($n=1$) to the continuum, the wave function of the outward-propagating photoelectron scatters off surrounding atoms. This interference produces an oscillatory fine structure which is characteristic of the chemical species of the absorber. Thus the oscillatory modulations of the cross section near the photoelectric absorption edge are unique fingerprints of the dust. Using high resolution X-ray spectra it is possible, in principle, to determine the properties of dust and complex molecules.\\
  In the X-rays we can observe absorption due to gas and dust in cold ($<10^3\ $K) and in hot ($ \geq 10^7\ $K) environments over a wide range of column densities $(N_{\text{H}} \sim 10^{20 - 24}\ \text{cm}^{-2})$. Thus, using the X-ray spectra we can also determine abundances in different ISM environments, thereby opening a window on the study of grain evolution and the cycle between diffuse and dense or dark clouds. Dust in diffuse regions along the Galactic plane has been modelled in the X-rays by \cite{Lee09, Costantini12, Pinto13, Valencic15} and \cite{Zeegers17} for several lines of sight. The best targets for this kind of study are low mass X-ray binaries (LMXB), which generally emit a featureless continuum X-ray spectrum. Therefore, only absorption features (lines and edges) from dust and from both neutral and possibly ionised gas are visible.\\
  
\noindent  
In this article, we present the K-edge XAFS laboratory data of various iron compounds measured at the European Synchrotron Radiation Facility (ESRF). These measurements are part of a large laboratory measurement campaign aimed at the characterisation of interstellar dust analogues \citep{Costantini13}. The investigated mineral set contains both crystalline and amorphous silicates, as well as iron sulfides, and is presented in detail in Section \ref{sec:sample}. Finally, for the analysis of synchrotron radiation based X-ray absorption spectroscopy, we discuss the prospect of studying the composition of iron grains in the ISM with the microcalorimeters onboard the future missions XARM (the successor of Hitomi) and Athena.\\
The paper is organised as follows. In Section \ref{sec:data} we describe the synchrotron measurement campaign and the data analysis. Section \ref{sec:esa} illustrates the entire procedure to calculate the final extinction cross sections taking into account both the absorption and the scattering phenomena. In Section \ref{sec:fek} we discuss the properties and the potential of the Fe K-edge to investigate the silicate dust grains. The iron K-edge will be studied in detail with the future X-ray missions presented in Section \ref{sec:sim}. Section \ref{sec:con} summarises the points discussed in this paper.

\section{Laboratory data analysis}
\label{sec:data}
\subsection{Sample}
\label{sec:sample}

According to infrared observations of $9.7$ and $19\ \mu$m features, the silicate interstellar dust mixture consists of an olivine (Mg$_{2-x}$Fe$_{x}$SiO$_4$) and pyroxene (Mg$_{1-x}$Fe$_{x}$SiO$_3$) stoichiometry \citep{Kemper04,Min07}. In order to reproduce laboratory analogues of astronomical silicates, our sample set contains both pyroxene and olivine silicate with variation in the Mg:Fe ratio. The values of $\text{Mg}/(\text{Fe}+\text{Mg})$ of our compounds range between $0.6$ for iron-rich silicates and $0.9$ for iron-poor silicates. Our choice of this variation depends on previous studies about the astronomical silicate composition, see e.g. \cite{Jager98, Kemper04, Min07, Olofsson09, Zeegers17}.\\ 
Iron is not only found in silicate structures. A likely possibility is that Fe exists in metallic form included in larger particles \citep{Costantini12, Jones13}. \cite{Kohler14} also infer a FeS inclusion in silicate grains. The bulk of the iron sulphide is formed either in the collapse phase of the molecular cloud or in the protoplanetary disc \citep{Keller02}. There is also evidence for FeS in a few planetary nebulae, see \cite{Hony02}. In order to investigate the Fe-bearing grains in the interstellar dust we included iron sulfide compounds (Fe$_{1-\text{x}}S$ with $x=0-0.2$) in our sample set.\\
In summary, we analysed a sample of six silicates and two iron sulfides (see Table \ref{tab:sample}). The sample set contains both amorphous and crystalline structures for two pyroxene compositions with different Mg:Fe ratios (samples 2, 3, 6, and 7). The amorphous samples were synthesised in the laboratory following the procedure described by \cite{Dorschner95}. The crystalline counterparts of these compounds were also artificially produced by slowly cooling the silicate mixture. The Si-bearing compounds present in the sample set correspond to the silicates already presented in detail by \cite{Zeegers17}. The two iron sulfides have different origins. The pyrrothite was synthesised in the laboratory at Astrophysikalisches Institut und Universit$\ddot{\text{a}}$ts-Sternwarte (AIU), whereas the troilite has a meteoritic origin.\\

\noindent
The absorbance of metallic Fe is well documented in the literature. Since the metallic iron was missing from our synchrotron sample set, we selected the data from the handbook of Exafs Materials\footnote{\url{http://exafsmaterials.com/ReferenceSpectra.html}}. In this work the energy range of the tabulated spectra includes the extended X-ray fine structures up to $1\ \text{keV}$ from the nominal edge energy. The X-ray spectrum of the metallic iron foil was recorded at synchrotron beamlines in the Stanford Synchrotron Radiation Laboratory (SSRL). The spectrum was taken in the transmission geometry with a spectrometer resolution of 0.5 eV in the region near the Fe absorption K-edge. 

%
%

\begin{table}
\caption{List of samples in our set with their relative chemical formulae. The nomenclature En(x)Fs(x-1) reveals the fraction of iron (or magnesium) included in the compound; $\lq$En' stands for enstatite (the magnesium end-member of the pyroxene silicate mineral series, MgSiO$_3$) and $\lq$Fs' for ferrosilite (the respective iron end-member, FeSiO$_3$).}  
\label{tab:sample}      
\centering                      
\begin{tabular}{c l l l}    
\hline\hline 
\# & Name & Chemical formula & State \\  
\hline               
 1 & Olivine\tablefootmark{a}      	& Mg$_{1.56}$Fe$_{0.4}$Si$_{0.91}$O$_4$  	& crystalline	\\
 2 & En60Fs40\tablefootmark{b} 		& Mg$_{0.6}$Fe$_{0.4}$SiO$_3$         		& amorphous	\\
 3 & En60Fs40\tablefootmark{b}   	& Mg$_{0.6}$Fe$_{0.4}$SiO$_3$         	 	& crystalline	\\
 4 & Troilite\tablefootmark{c}   	& FeS 	                              		& crystalline	\\
 5 & Hypersthene\tablefootmark{d}      	& Mg$_{1.502}$Fe$_{0.498}$Si$_2$O$_6$  		& crystalline	\\
 6 & En90Fs10\tablefootmark{b} 		& Mg$_{0.9}$Fe$_{0.1}$SiO$_3$         	 	& crystalline	\\
 7 & En90Fs10\tablefootmark{b}   	& Mg$_{0.9}$Fe$_{0.1}$SiO$_3$         		& amorphous	\\
 8 & Pyrrhotite\tablefootmark{b}       	& Fe$_{0.875}$S                         	& crystalline	\\

\hline                        
\end{tabular}
\tablefoottext{a}{origin: Sri Lanka}
\tablefoottext{b}{synthesised in laboratories at AIU Jena and Osaka University}
\tablefoottext{c}{meteoritic origin}
\tablefoottext{d}{origin: Paul Island, Labrador}
\end{table}

\subsection{Synchrotron measurements}
The data presented in this paper (except metallic iron, see above) were measured at the European Synchrotron Radiation Facility (ESRF) in Grenoble, France. We used the Dutch-Belgian beamline (DUBBLE\footnote{\url{http://www.esrf.eu/UsersAndScience/Experiments/CRG/BM26}}). The beamline BM26A is specialised in X-ray absorption fine structure spectroscopy \citep{Nikitenko08}. In essence, the synchrotron radiation emitted by a bending magnet source is monochromatised using an energy tuneable monochromator able to select a narrow energy range $(\frac{\Delta \lambda}{\lambda}=3\times 10^{-4})$. By rotating the monochromator in discrete steps it is possible to scan the absorption spectrum around the Fe K-edge over the range $6.8 - 8.2$~keV with an accuracy of 1\% and a resolution of 0.3~eV. During an energy scan the beam position systematically changes, and therefore sample homogeneity is required. For this reason the finely powdered samples were diluted uniformly in a boron nitride matrix used as a spacer and sticker. Afterwards the mixed material was pressed into pellets. The final thickness of the sample was enough to obtain a significant edge jump ($>0.1$ of the absorption length). For $13\ $mm diameter pellet sample we used $20\ $mg of material and $100\ $mg of boron nitride. The contrast in the XAFS features increases as the amplitude of thermal vibrations of the atoms decreases, so we cooled our samples to 90~K. To reach this temperature the pellets were placed into a cryostat. Several (4-6) X-ray absorption scans in transmission were measured on each sample to ensure the reproducibility of the spectra and to obtain a high signal-to-noise ratio. In addition to the sample observations, a baseline pellet of pure boron nitride without any sample inserted was measured to obtain the characteristics of the bonding material. For each measurement the incident and the transmitted beam intensities were measured using three ionisation chambers mounted in series for simultaneous measurements on the sample.\\

\subsection{Analysis of the laboratory data}
\begin{figure}
   \centering
   \includegraphics[width=\hsize]{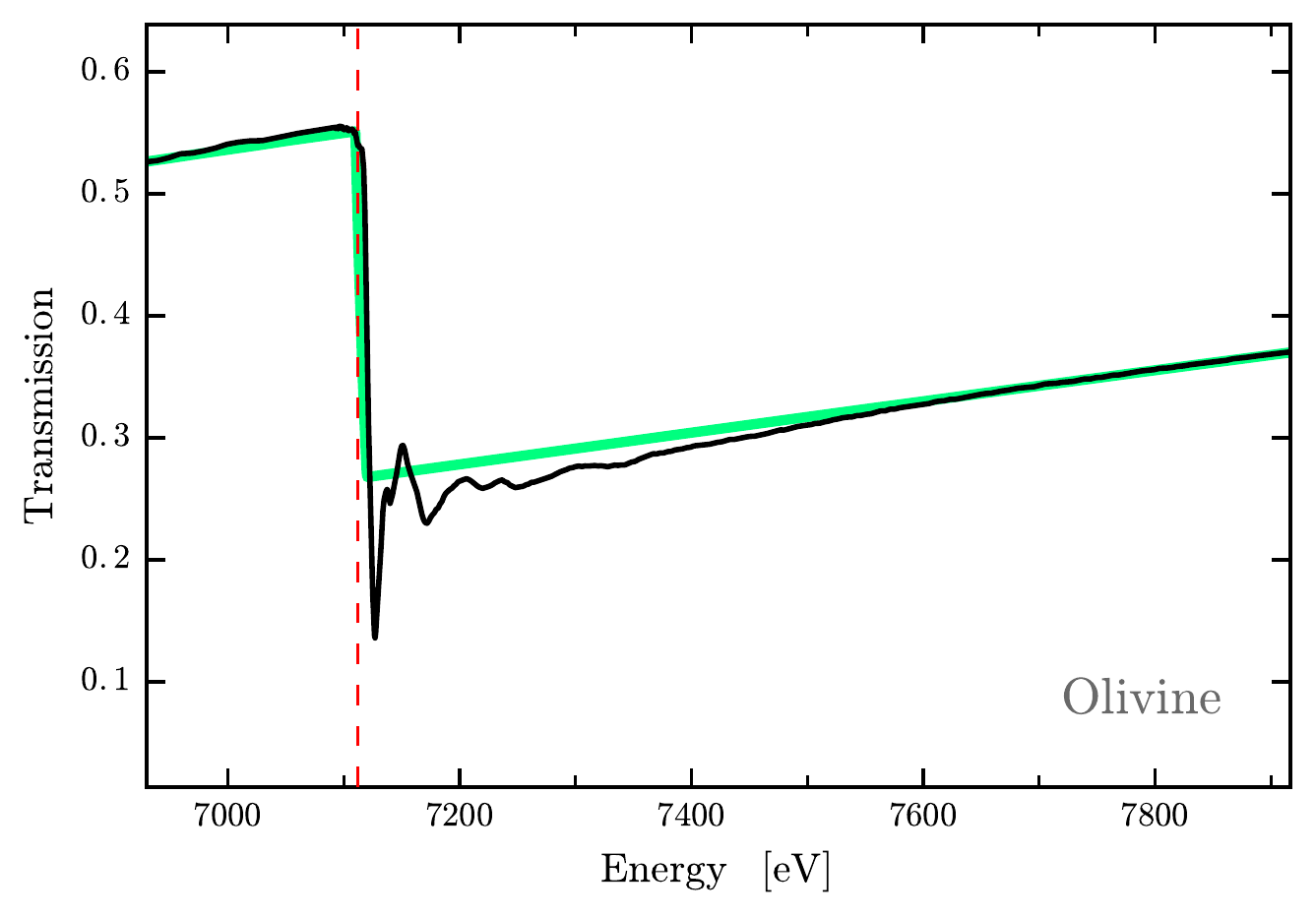}
      \caption{Transmission signal around the Fe K-edge measured at the Dutch-Belgian beamline (DUBBLE) at the European Synchrotron Radiation Facility (black solid line). The sample used in this measurement is olivine. The vertical dashed red line indicates the energy of the Fe K-edge ($7112\ $eV). The green solid line represents the \cite{Henke93} data table used to fit the laboratory measurement. 
              }
         \label{fig:transmittance}
   \end{figure}

When an X-ray photon with energy ranging from $\sim 100$\,eV up to hundred keV interacts with material, it can be transmitted, or it can interact with the medium by photoabsorption or by coherent scattering at the electron shell of the atoms in the material. The amount of light that is transmitted depends on the original energy of the X-rays, on the optical path length, and on the properties of the material. This quantity is called transmittance $T$ and is defined as the ratio of the transmitted $I$ and the incident light $I_0:$
\begin{equation}
T=I/I_0\ .
\label{eq:transmittance}
\end{equation}
Figure \ref{fig:transmittance} shows the relative transmittance obtained from the synchrotron measurements of olivine (sample 1 in Table \ref{tab:sample}).\\ 
The normalisation was performed using a standard procedure \citep{Koningsberger88}, regressing a second-order polynomial to the region before the pre-edge. We subtracted it from the entire spectrum to eliminate any instrumental background and absorption from other edges. A third-order polynomial was regressed to a region beyond the absorption edge. The value of the post-edge polynomial extrapolated back to the edge energy was used as the normalisation constant. The attenuation of X-rays travelling through a material of thickness $x$ (in $\mu$m) is also given by the Beer-Lambert law, which states that the intensity decreases exponentially with depth in material according to 
\begin{equation}
I(E)=I_0 e^{-\mu(E)x}\ ,
\label{eq:lambert}
\end{equation}
where $\mu(E)$ is the attenuation coefficient in $\mu$m$^{-1}$. If we ignore reflection (at X-rays energies, the contribution of reflection becomes very small) we can equate the attenuation coefficient with the absorption coefficient. Therefore, the characteristic attenuation coefficient for a compound is approximately given by
\begin{equation}
\mu(E) \approx \sum_{\text{i}} \rho_{\text{i}} \sigma_{\text{i}} = \rho_{\text{M}} \sum_{\text{i}} \frac{m_{\text{i}}}{M}\sigma_{\text{i}}\ ,
\end{equation}
where $\rho_{\text{M}}$ is the mass density of the compound, $m_{\text{i}}/M$ is the fractional mass contribution of element $i$, and $\sigma_{\text{i}}$ is the respective mass normalised absorption cross section. In our samples we keep the total thickness $x$ below $2-3$ absorption lengths to minimise thickness effects which may reduce the EXAFS amplitude \citep{Parratt57, Bunker10}. In order to determine $\mu$ from our synchrotron measurements we use the relation with transmittance obtained by setting Equation \ref{eq:transmittance} and \ref{eq:lambert} equal to each other:
\begin{equation}
\mu(E) = \frac{-\ln{T}}{x}\ .
\end{equation}
The experimental $x$ is not known a priori since the sample is pressed in pellets and diluted. In order to evaluate the thickness we compare our laboratory transmittance signal with the tabulates values provided by the Center for X-ray Optics at Lawrence Berkeley National laboratory\footnote{\url{http://www.cxro.lbl.gov/}}. Using the $\chi^2$ test we calculate the best value for the thickness, the only free parameter. To achieve this result we fit the pre- and post-edge regions. Obtaining the value of the attenuation coefficient is important in order to calculate the optical constants ($n$, $k$) of the refractive index of the material.

\section{Extinction, scattering, and absorption cross sections}
\label{sec:esa}

The refractive index is a complex and dimensionless number that describes how light propagates through a specific material. In the following paragraph we describe the optical constants which identify the refractive index. We summarise the different definitions present in the literature. Subsequently, we present the Kramers-Kronig \citep{Kronig26, Kramers27} relations and the anomalous diffraction theory (ADT, \cite{Hulst57}) that we use to obtain the extinction cross section.

\subsection{Optical constants}
\begin{figure}
   \centering
   \includegraphics[width=\hsize]{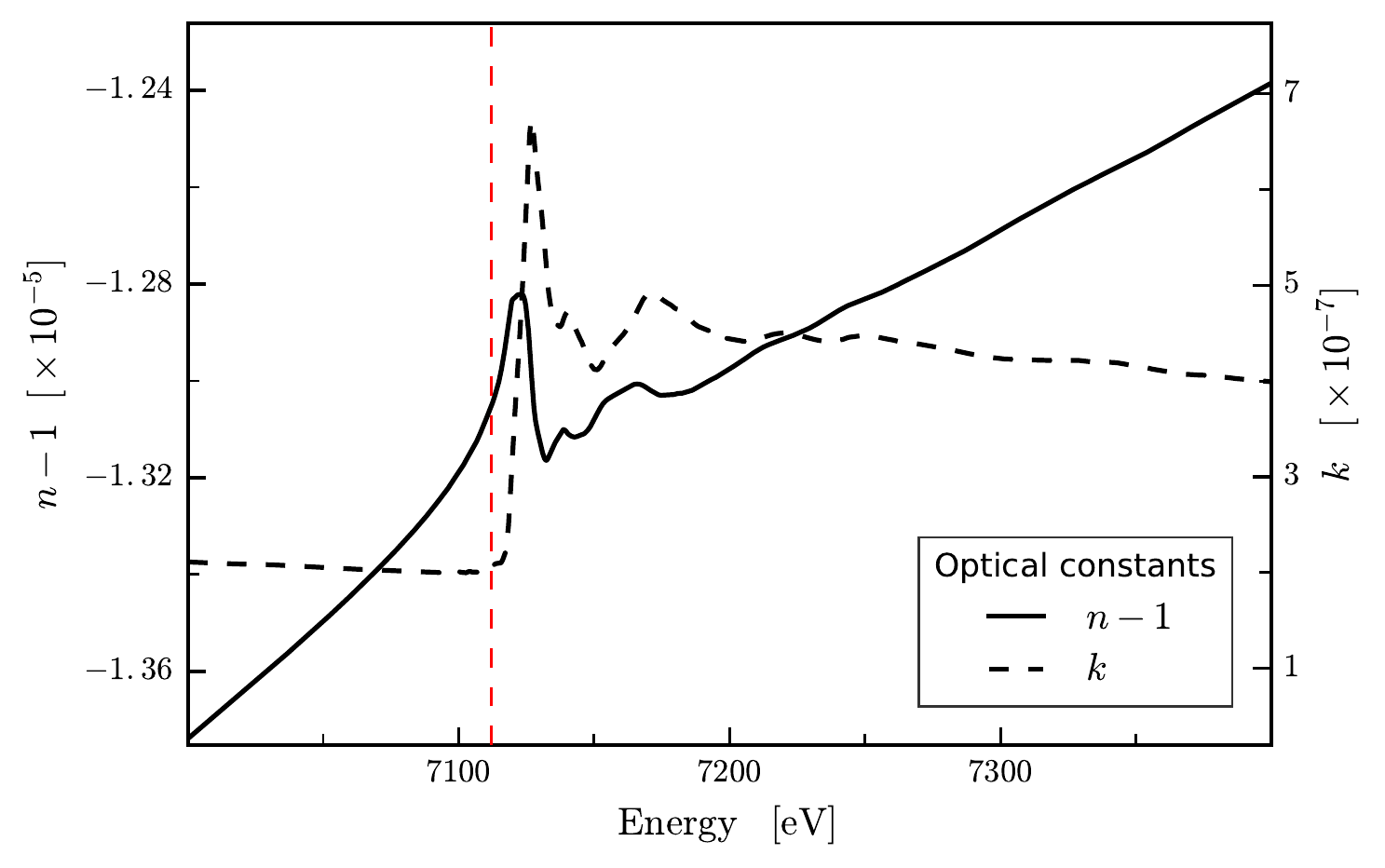}
      \caption{Real (solid line) and imaginary (dashed line) part of the refractive index for olivine. The optical constant $k$ is derived from synchrotron measurement using Equation \ref{eq:k}. The  Kramers-Kronig relations (see Equation \ref{eq:kramers}) were used to calculate $n$.
              }
         \label{fig:optical}
   \end{figure}

The refractive complex index is generally defined as
\begin{equation}
\centering
m= n+ik\ ,
\label{equation:ind_refractive}
\end{equation}
where $n$ (or Re($m$)) describes the dispersive behaviour and $k$ (or Im($m$)) the corresponding absorption depending on the energy of the incident light. For X-rays, the complex refractive index deviates only slightly from unity and usually the real part is smaller than 1. It is therefore commonly written as $|m-1|$. We show an example in Figure \ref{fig:optical} for olivine. The refractive index has an analogue notation used in the literature \citep[e.g.][]{Henke93}
\begin{equation}
\centering
n_{\text{r}}=1-\delta +i\beta\ ,
\end{equation}
where $n_{\text{r}}$ corresponds to $m$ in Equation(\ref{equation:ind_refractive}) and $\delta$ and $\beta$ are the optical constants. Other notations use the complex dielectric functions, $\varepsilon_1$ and $\varepsilon_2$ \citep{Landau60, Draine03} defined as
\begin{equation}
\varepsilon= \varepsilon_1 + i\varepsilon_2= m^2\ .
\end{equation}
It is sometimes useful to make explicit their relation to the optical constants $n$ and $k$:
\begin{align}
&n=\sqrt{\frac{|\varepsilon |+\varepsilon_1}{2}}\ ,\\
&k=\sqrt{\frac{|\varepsilon |-\varepsilon_1}{2}}\ .
\end{align}
Finally, another common notation uses the atomic scattering factors, $f_1$ and $f_2$ \citep{Henke93}   
\begin{equation}
f(E)= f_1(E) + if_2(E)= \frac{2\pi A}{\rho N_\text{A} r_0}\cdot\frac{1}{\lambda^2}\cdot(m-1)\ .
\end{equation}
They are related to the optical constants as
\begin{align}
&n(E)=1-\frac{ \rho N_{\text{A}} r_0}{2\pi A}\lambda^2\cdot f_1(E)\ ,\\
&k(E)=\frac{ \rho N_{\text{A}} r_0}{2\pi A}\lambda^2\cdot f_2(E)\ ,
\label{eq:conversion_k_f2}
\end{align}
where $N_{\text{A}}$ is the Avogadro's number, $A$ is the atomic mass number of the compound, $\rho$ is the density in $\text{g}/\text{cm}^3$, $r_0$ is the classical electron radius, and $\lambda$ is the wavelength of the incident X-ray.

\subsection{Kramers-Kronig relation}
\label{sec:kk}
The imaginary, absorptive part of the refractive index ($k$) can be easily determined via laboratory experiments. Indeed, there is a relation between the absorptive optical constant $k$ and the linear absorption coefficient $\mu$: 
\begin{equation}
k=\frac{\mu\lambda}{4\pi}\ .
\label{eq:k}
\end{equation}
In this way we can determine $k$ as a function of energy (or wavelength) in detail around the edge \citep[see also][]{Zeegers17}. The real dispersive part, on the other hand, is more difficult to measure. Fortunately, the real and imaginary parts are related to each other by the Kramers-Kronig relations and therefore the refractory index can be determined from measurements of only the imaginary part. To calculate the real part of the Henke-Gullikson scattering factor $f$, we use the Kramers-Kronig relation expressed in the equation \citep{Gullikson01}
\begin{equation}
f_1(E)=Z^* -\frac{2}{\pi}P \int^{\infty}_{0} \frac{xf_2(x)}{x^2-E^2}\,dx\ ,
\label{eq:kramers}
\end{equation}
where $E$ in this case is the photon energy at which we wish to calculate the real part of the atomic scattering factor, $Z^*$ is the relativistic correction, and $P$ denotes the Cauchy principal value \citep{Landau60}. Equation (\ref{eq:kramers}) is problematic in two ways. First, the integral spans from zero to infinity, meaning that in order to calculate $f_1$ over a small energy range, the value of $f_2$ is required over a virtually infinite energy range. The second issue is that the integral contains a discontinuity at $x=E$ and therefore must be considered as a Cauchy principal value. The Cauchy principal value is central to many calculations and a number of resolving strategies have been applied to the problem, such as the fast Fourier transform \citep[FFT,][]{Bruzzoni02}, the Euler-MacLaurin \citep{Cross98}, and the Chebyshev methods \citep{Hasagawa91}. \cite{Henke93} demonstrated a simple method involving a manual piecewise polynomial representation of spectral data. This way of evaluating the integrals is accurate but time-consuming. \cite{Watts14} introduces an useful extension of the \cite{Henke93} approach. It allows computer automation and it can provide computationally fast calculations whose accuracy is limited only by the representation of the absorption spectrum by a piecewise set of Laurent polynomials. This algorithm calculates the relativistic correction following the \cite{Henke93} approach,
\begin{equation}
Z^*= \sum_{q} n_{\text{q}}\,\Bigg(Z_{\text{q}}-\bigg(\frac{Z_{\text{q}}}{82.5}\bigg)^{2.37}\Bigg)\ ,
\end{equation}
where $n_{\text{q}}$ and $Z_{\text{q}}$ represent the number density of atoms and the atomic number of the $q$th element in the material, respectively. This method for calculating the real part of the scattering factors from the imaginary has several benefits: \textit{i)} it does not require a homogeneous binning of data points, \textit{ii)} a full energy range from zero to infinity is not necessary, \textit{iii)} it can process broad sections of spectrum at once, while maintaining accuracy and precision. Thus, it is suitable for our laboratory transmission, which is measured only for a limited range of frequency around the absorption edge. The method has been implemented as a Python-based library\footnote{\url{https://bitbucket.org/benajamin/kkcalc}} \citep{Watts14}. 

\subsection{Anomalous diffraction theory}
\begin{figure}
   \centering
   \includegraphics[width=\hsize]{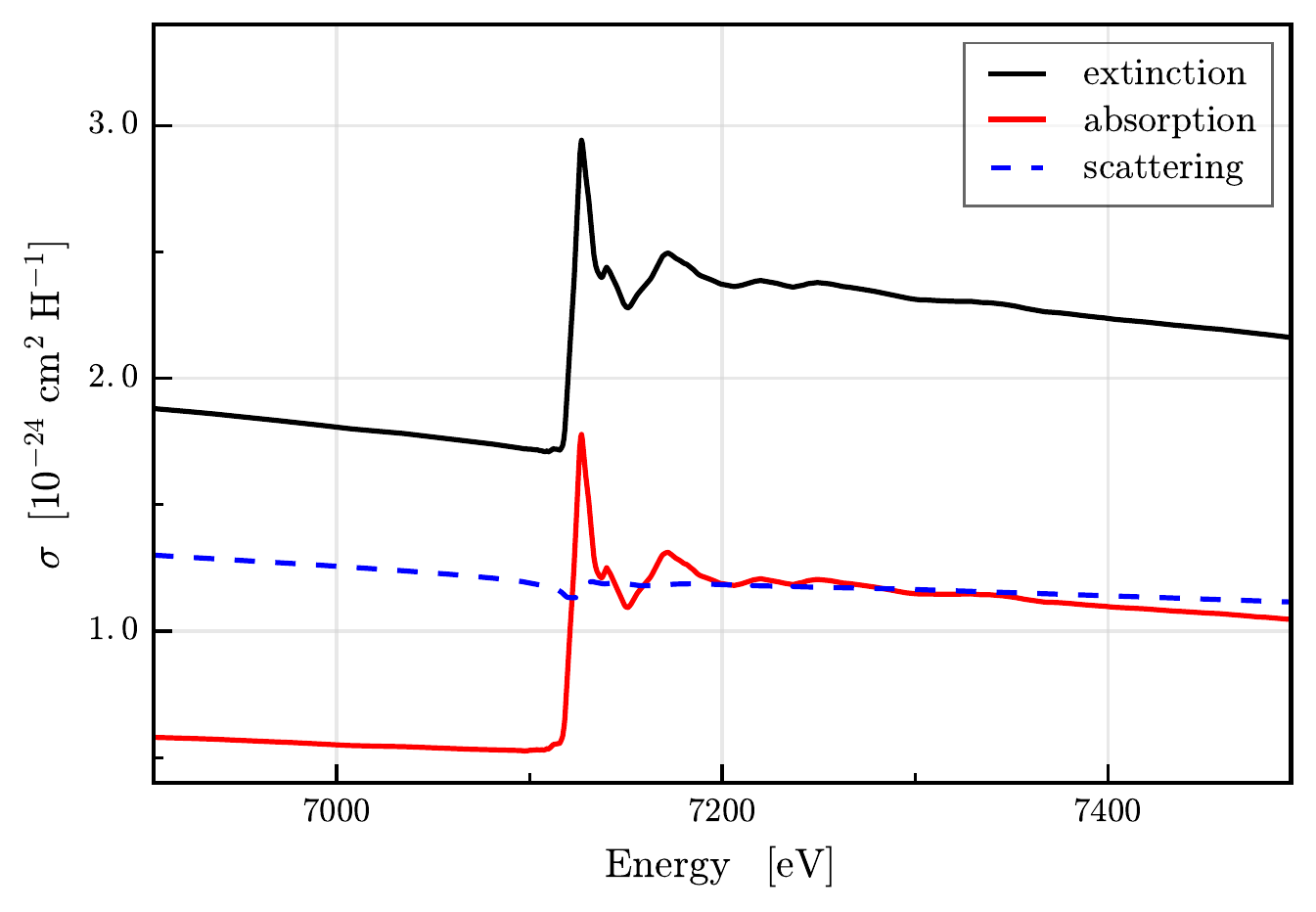}
      \caption{X-ray extinction (solid black line), absorption (solid red line), and scattering (dashed blue line) cross sections per H nucleon for the \cite{Mathis77} dust model ($0.005 \le a \le 0.25\ \mu\rm{m}$). The scattering cross section contributes significantly to the extinction, but it does not significantly modify the absorption features in the post-edge. This plot is characteristic of olivine. 
              }
         \label{fig:cross}
   \end{figure}

To calculate the extinction cross section we used the anomalous diffraction theory\footnote{Not to be confused with anomalous diffraction (also known in X-ray crystallography as anomalous scattering).}, an algorithm for computing absorption and scattering by the dust grains. The merits of ADT are its simplicity in concept and efficiency in numerical computation. Moreover it is applicable to grains of arbitrary geometry that are larger than the incident wavelength. The approximations used in the derivation of ADT require that $|m-1|\ll 1$, and $x=2\pi a / \lambda\gg1$, where $a$ is the effective radius of the grains. In other words the particle has to be optically soft (i.e. the refractive index has to be close to unity) and larger than the wavelength of incident light. \cite{Hoffman16} show that the ADT approximation fits well for our range of wavelengths (around the Fe K-edge at $7.112\ $keV) and grain radius size (up to 1 $\mu$m). Because $x\gg 1 $, the concept of independent rays of light passing through the grains is valid. And, because $|m-1|\ll1$, refraction and reflection effects are small and they can be ignored.\\
The ADT calculates the efficiency for absorption, scattering, and extinction efficiencies per grain size at each energy of interest. We use the following equations expressed in terms of $n$, $k$, and $x$ (see \cite{Hulst57} for a different notation),
\begin{align}
&Q_{\text{ext}}= 2+4\Bigg[\frac{\cos(2\beta) - e^{-2xk}\cos(2x(n-1)-2\beta)}{4x^2((n-1)^2+k^2)}+\\
&\qquad\qquad+\frac{2x\sqrt{(n-1)^2+k^2}\cdot\sin(2x(n-1)-\beta)}{4x^2((n-1)^2+k^2)}\Bigg]\ , \\
&Q_{\text{abs}}= 1+\frac{e^{-4xk}}{2xk}+\frac{1}{2}\cdot\frac{e^{-4xk}-1}{(2xk)^2}\ ,\\
&Q_{\text{sca}}=Q_{\text{ext}}-Q_{\text{abs}}\ ,
\end{align}
where $\beta = \arccos\Big((n-1)/\sqrt{(n-1)^2+k^2}\Big)$. To obtain the total cross sections per wavelength unit ($ C = \pi a^2 \cdot Q$), we need to integrate over the particle size distribution. 
We use for the moment the Mathis-Rumpl-Nordsieck (MRN) grain size distribution \citep{Mathis77} with a grain size interval of $(0.005\leq a \leq 0.25)~\mu$m,
\begin{equation}
n(a)\, da=A\cdot a^{-3.5}\, da\ ,
\end{equation}
where $a$ is the particle size, $n(a)$ is the number of grains, and $A$ is the normalisation constant, which depends on the type of dust. The value of the constant for each compound can be determined following the method described by \cite{Mauche86}. In this work they use the Rayleigh-Gans approximations to calculate $A$. This approximation requires that $
\big( \frac{4\pi a}{\lambda}\big)\;|m-1|\ll1\ .$
Consequently, we cannot use this approximation in our analysis since for the grain size distribution that we take into account the mean energy of the sources should be less than $2\ $keV. Therefore, we apply the ADT approximation to evaluate the normalisation constant (see Zeegers et al. 2017, in prep). In Figure \ref{fig:cross} we plot the absolute cross sections for olivine, calculated with the procedure described above.\\
The last step, in order to compare our laboratory measurements with astronomical data, is to implement the extinction profiles in the SPEX fitting code \citep{Kaastra96}. Our measurements were added to the already existing AMOL model. Now this model assumes Verner absorption curves for the pre- and post-edge range \citep{Verner96}.

\section{Fe K-edge properties}
\label{sec:fek}
The iron K absorption edge is sensitive to particular physical and chemical characteristics of the dust grains. In this section we present the characteristics of the Fe K-edge and how the edge can help to study and investigate several properties of the grains.

\begin{figure}
   \centering
   \includegraphics[width=\hsize]{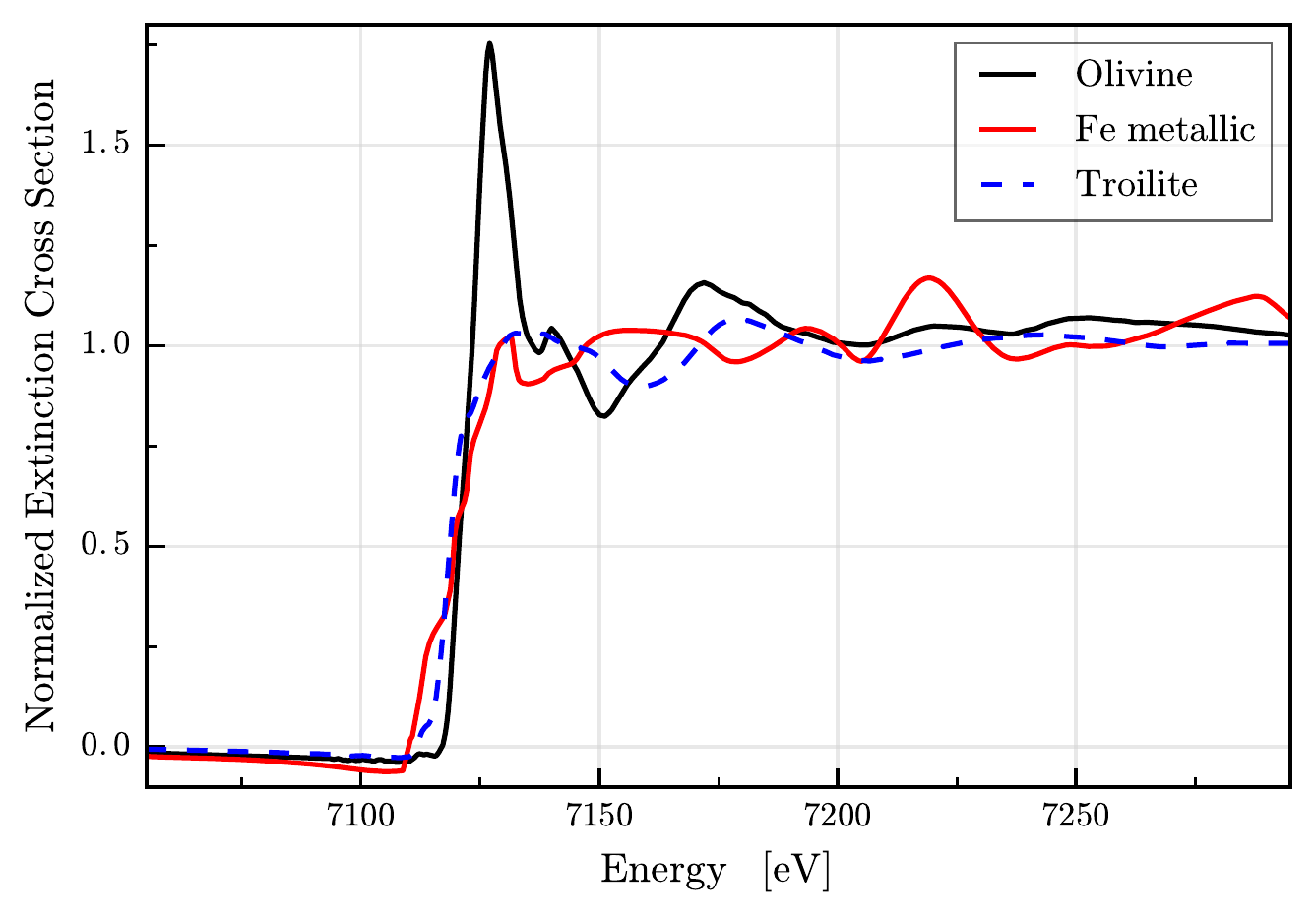}
      \caption{Extinction cross section in normalised units to compare the different chemical species presented here: silicate (olivine, black solid line), iron sulfide (troilite, blue dashed line), and metallic Fe (red solid line). The X-ray absorption features are different for these three compounds. In particular the silicate is characterised by the large peak just after the Fe K-edge located at $\rm{E}\sim7130\ $eV. 
              }
         \label{fig:chemistry}
   \end{figure}

\paragraph{\it{Chemistry \ -}}
As already shown in the introduction, the modulation of the post edge region is in principle the fingerprint for each specific compound.
In Figure \ref{fig:chemistry} we show the extinction cross section of three representative compounds in our sample set: a silicate (olivine, solid black line), an iron sulfide (troilite, blue dashed line), and Fe in the metallic form (red solid line). The silicates especially show characteristic X-ray absorption near edge structures (XANES) compared to the other compounds (see Appendix \ref{app:samples}). Indeed the extinction cross section of olivine displays a higher peak right after the rising edge, at $\sim 7130$\ eV. Metallic Fe and troilite do not show large features in the post-edge region. The iron sulfides are characterised by two wide peaks at $\sim 7135\ $eV and $\sim 7175\ $eV. Instead, the absorption cross section of the metallic iron presents a distinctive peak at higher energies ($\sim 7270\ $eV). All these features are useful to distinguish the compounds that contribute to the dust extinction cross section.

\begin{figure}
   \centering
   \includegraphics[width=\hsize]{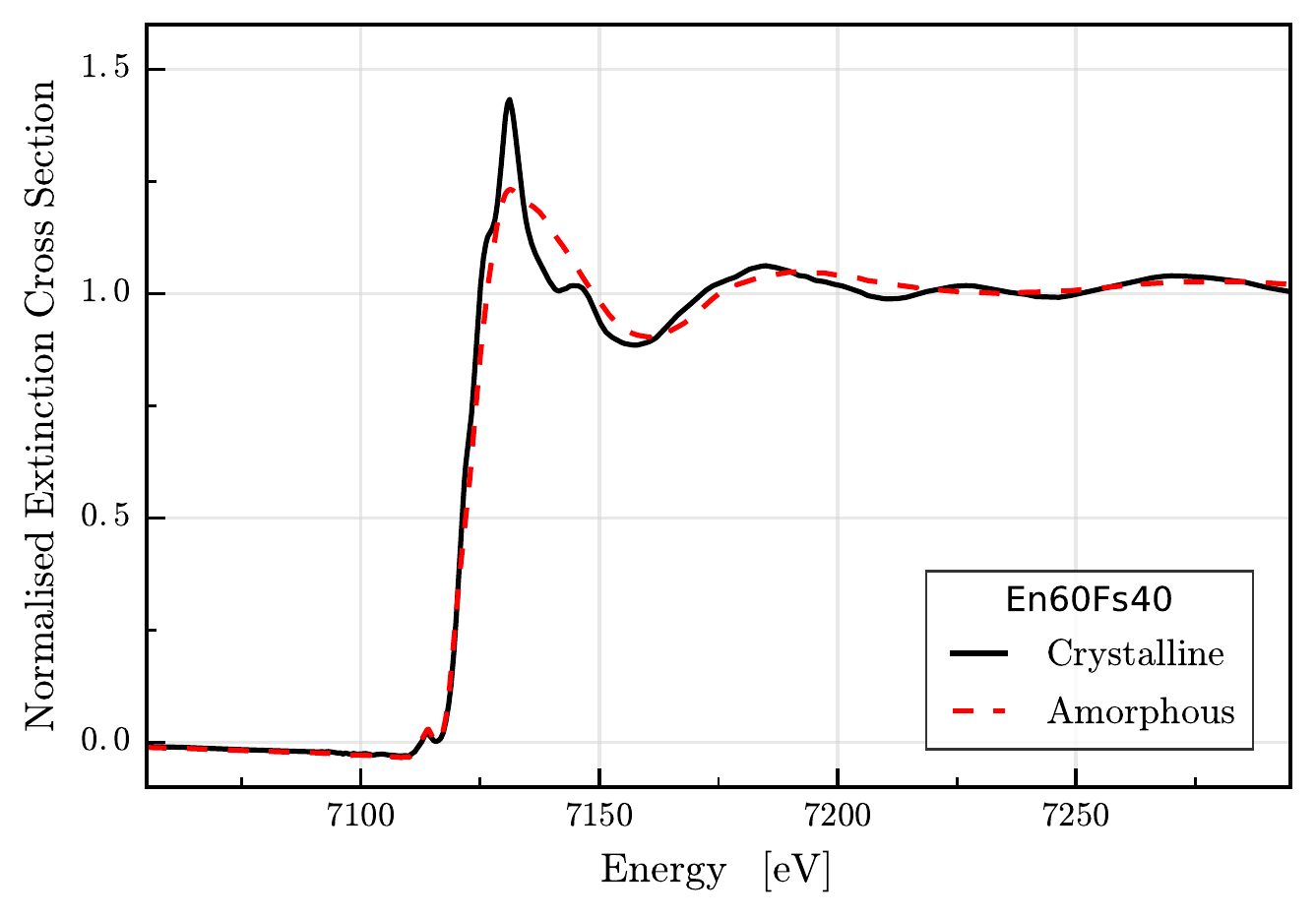}
      \caption{Extinction cross section in arbitrary units for pyroxene En60Fs40 in crystalline state (black and solid line) and in amorphous state (red and dashed line). 
              }
         \label{fig:crystalline}
   \end{figure}

\paragraph{\it{Glassiness \ -}}
The absorption cross section is also sensitive to the crystalline order of the compound. In the pure crystalline form the atoms have fixed position and form a periodic arrangement. Instead, an amorphous material is a solid that lacks the long-range crystalline order and thus the absorption cross section shows a different pattern of peaks. Figure \ref{fig:crystalline} compares the amorphous form (red dashed line) to the crystalline state (black solid line) for pyroxene En60Fs40 (see Table \ref{tab:sample}). The extinction cross section of the glass configuration is smoother and does not show any secondary peak near the edge. The characteristic peak of silicate at $\sim7130\ $eV (see the previous paragraph) is smaller and wider than the feature displayed in the crystalline form.

\paragraph{\it{Grain size\ -}}

   \begin{figure}
   \centering
   \includegraphics[width=\hsize]{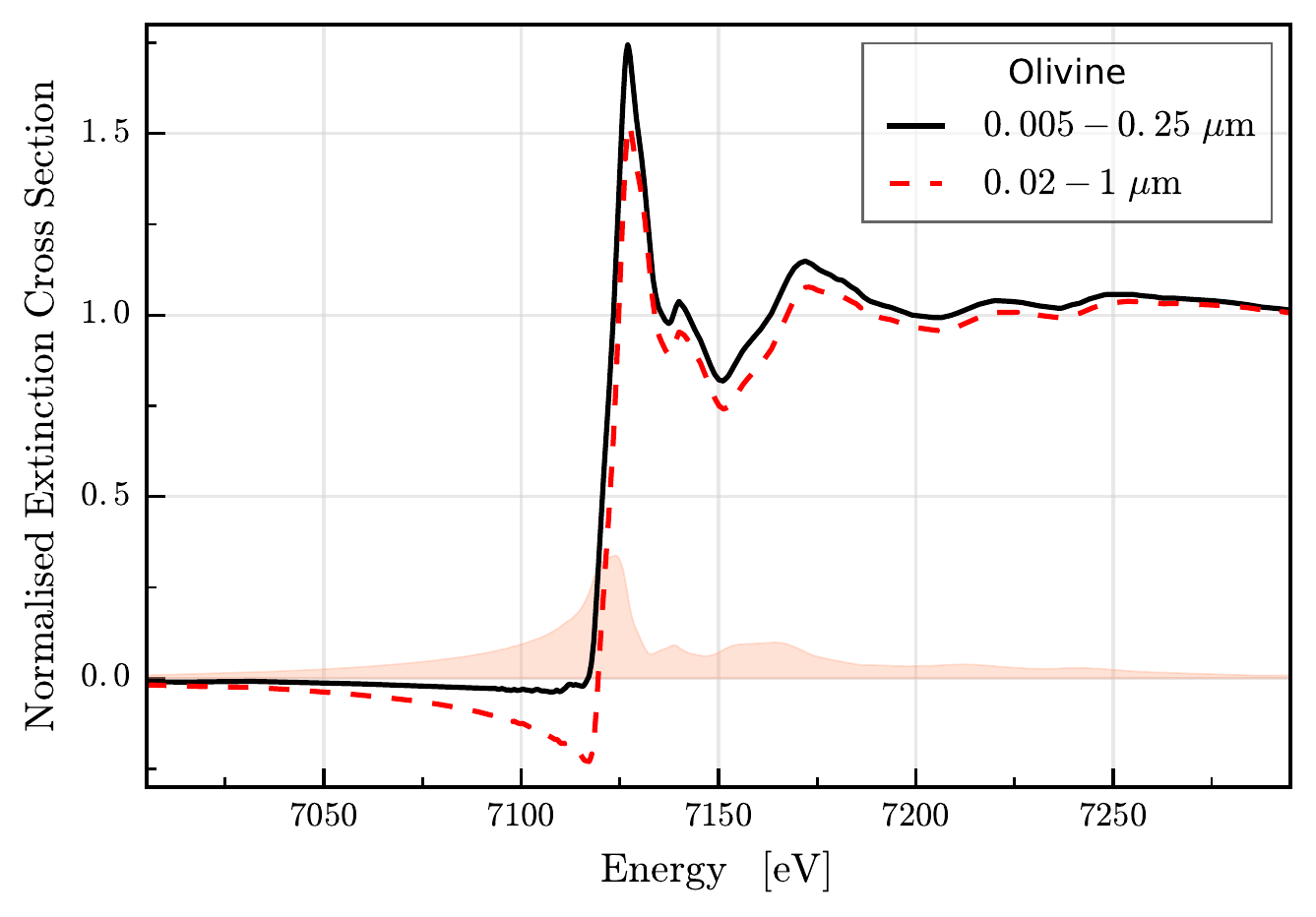}
      \caption{Normalised extinction cross sections for olivine calculated using two different grain size distributions. The solid black line represents the standard MRN grain size ranging between $0.005-0.25\ \mu$m. The dashed red line delineates the larger grain size spanning between $0.02-1\ \mu$m by \cite{Draine09}. The orange shadow in the bottom highlights the shape difference between the two extinction cross sections.    
              }        
   \label{fig:large_grains}
   \end{figure}
  
   \begin{figure*}
   \centering
   \includegraphics[width=.49\linewidth]{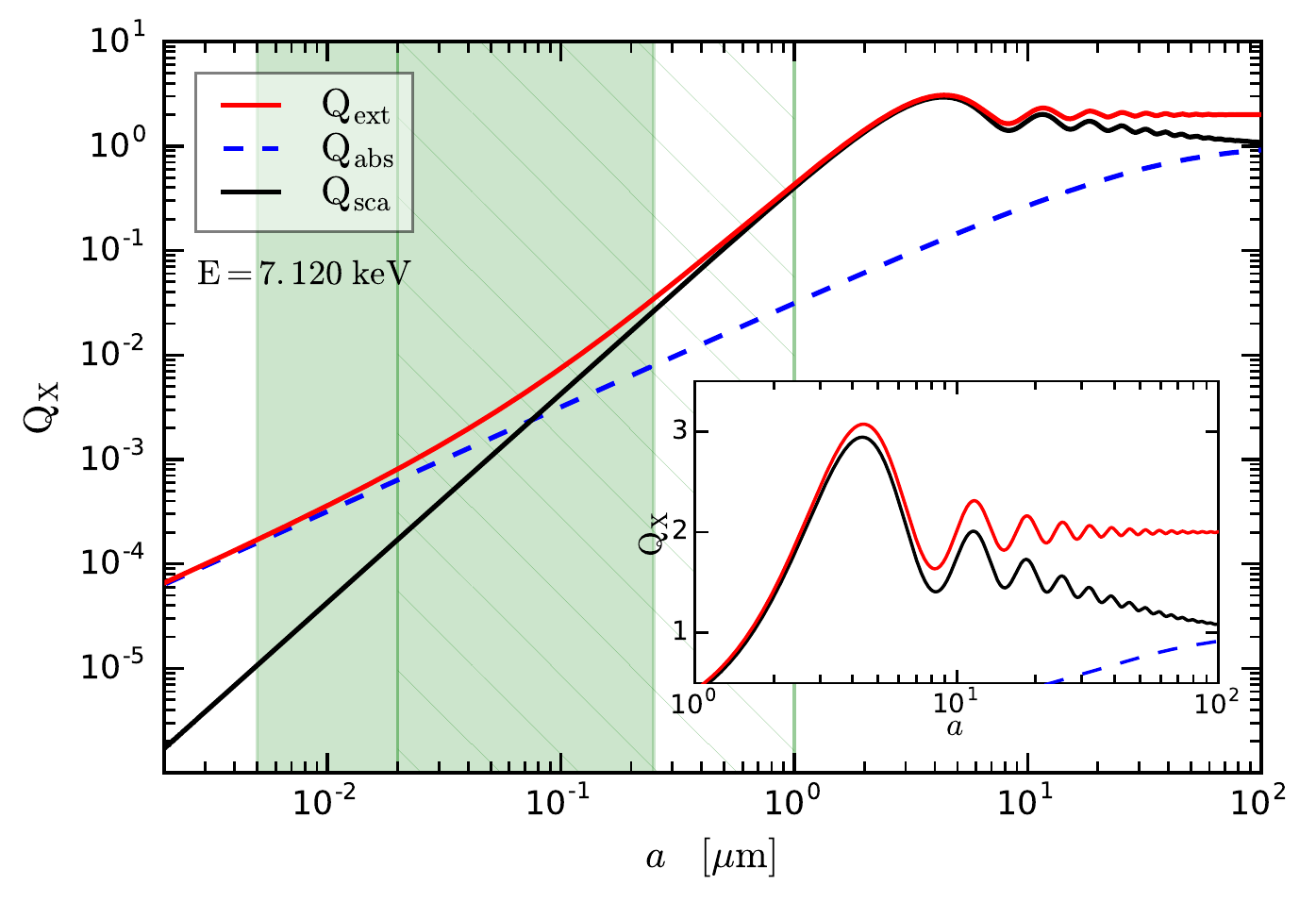}
   \includegraphics[width=.49\linewidth]{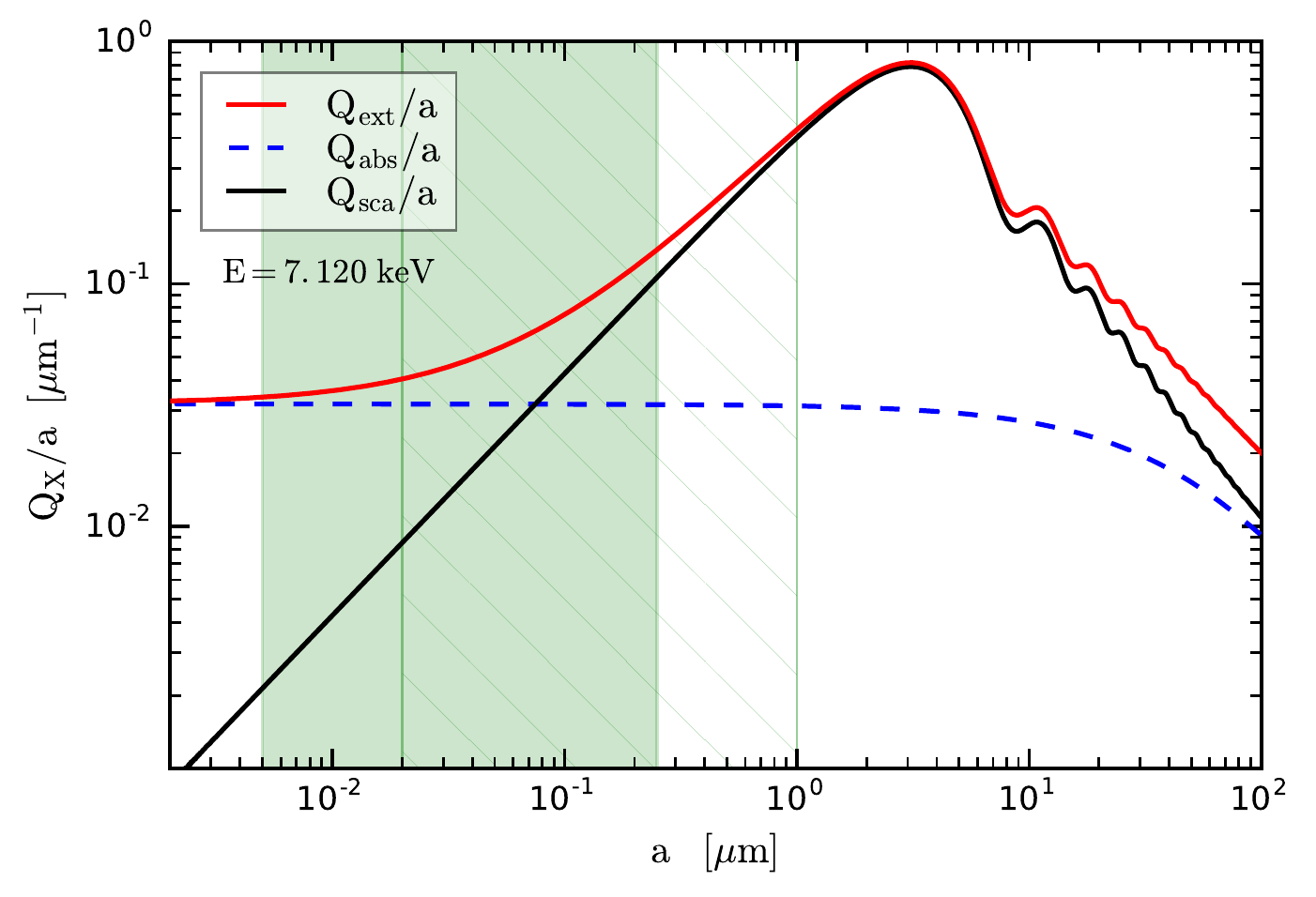}
      \caption{\emph{Left}: Comparison of the calculated absorption (dashed blue line), scattering (solid black line), and extinction (solid red line) efficiencies at $h\nu = 7120\ $eV (near Fe K-edge) for olivine. The two green regions indicate the size range investigated in this work: in particular ($0.005\le \rm{a} \le0.25)\ \mu$m (filled green) and $(0.02\le \rm{a} \le1)\ \mu$m (hatched green). \emph{Right}: Comparison of the calculated absorption, scattering, and extinction efficiencies per unit of radii as a function of the grain size. The absorption Q-factor per grain size remains roughly constant along the two size ranges analysed in this work (the green areas). Instead, the scattering Q-factor per radii unit increases, achieving the maximum value at $\sim3\ \mu\rm{m}$. 
              }
              
         \label{fig:sext_vs_a}
   \end{figure*}  
   
The scattering cross section is sensitive to the range in sizes ($a_- - a_+$) of the constituents. In general, the scattering cross section may potentially modulate the amplitude, the slope, and the features close to the absorption edges. To investigate the effect of the particle size distribution we integrate the scattering efficiency over different size ranges. We change the lower ($a_-$) and upper ($a_+$) cut-off of the MRN distribution and consequently the average grain size \citep{Mauche86}. \cite{Draine09} investigated particles between $a_-=0.02\ \mu$m and $a_+=1\ \mu$m for a model of spherical silicates grains. In order to study the enhancement of the scattering features around the iron K-edge we introduce this large particle size range with an average grain size $\bar{a}\sim 0.6\ \mu$m. The effect of the size distribution change is shown in Figure \ref{fig:large_grains}. The olivine Fe K-edge with the standard MRN distribution with particle sizes of $0.005-0.25\ \mu$m is shown in black (solid line). In red (dashed line) the same edge is shown, but now with a MRN size distribution that has a particle range of $0.02-1\ \mu$m. Here we compare the shape of the two extinction cross sections (normalised for this purpose). The main difference is the peak, due to the scattering cross section, right before the edge. This scattering feature can potentially be used to investigate the mean range size of the interstellar dust.\\
Figure \ref{fig:sext_vs_a} gives a general overview on the efficiency as a function of grain size for the energies investigated in this paper. We show a comparison among the calculated absorption, scattering, and extinction efficiencies at a fixed energy ($E = 7120\ $eV). In particular the scattering becomes more efficient for larger grains and dominates the extinction process for radii larger than $0.2\ \mu\rm{m}$. The scattering efficiency (and thus the extinction efficiency) continues to increase up to $\sim3\ \mu\rm{m}$ where it reaches the maximum value, as highlighted in the right panel of Figure \ref{fig:sext_vs_a}, which shows the Q-factors per unit radius. The scattering efficiency also depends on the energy of the incident light: for higher energies the peak of the scattering efficiency is shifted to larger grain radii (see Zeegers et al. 2017 in prep.). This peak is followed by smaller bumps due to the resonance of the diffraction \citep[see][]{Berg10}. From the trend of $Q_{\text{ext}}$, larger grains ($>1\ \mu$m) seem more efficient at extinguishing X-rays, but at the same time they are less abundant. However, we do not expect a significant grain population with radii larger than $1\ \mu$m even in the molecular core, as near- and mid-infrared observations show \citep{Andersen13}.

\paragraph{\it{Size distribution model and porosity\ -}}
\cite{Hoffman16} have studied in detail the X-ray extinction for astrosilicate grains \citep{Draine03}. They find that the Fe K-edge shape is not significantly affected by the choice of the dust size distribution tested \cite[namely MRN and][]{Weingartner01}. \cite{Hoffman16} also show that the geometry and the porosity level of the grains do not alter the fine structure of the absorption Fe K-edge. Thus, in the study of the iron K-edge porosity, both the effect of the porosity and dust size distribution can be at first order ignored. 

\section{Simulations}
\label{sec:sim}

\subsection{Present and future X-ray missions}

The laboratory measurements cannot be applied to astronomical data from the current X-ray missions because of insufficient energy resolution at $7.1\ $keV. In the following, we explore the prospects of observing the Fe K-edge with the resolution provide by the X-ray Astronomy Recovery Mission\footnote{The recovery mission XARM will have the same technical characteristic of the previous mission ASTRO-H specially for the microcalorimeter.\\ \url{http://astro-h.isas.jaxa.jp/en/}} (XARM, successor of Hitomi) and the Athena\footnote{see \cite{Nandra13} for an overview.} mission (see Table \ref{tab:xraymission}). The high energy resolution in the Fe K-edge will be achieved by using microcalorimeters.\\
With the Athena satellite it will be easier to probe the dense interstellar regions for the iron component of the dust grains. In particular, it will be possible to obtain direct measurements of the Fe abundance and inclusion in dust in dense regions only accessible through the Fe K-edge. The large effective area of the X-ray spectrometer X-IFU allows us to obtain high quality spectra not only of the brightest X-ray sources, but also of a fainter source population.\\
%
%

\begin{table}
\caption{Principal characteristics of the X-ray spectrometer of Chandra-HETG, XARM-SXS, and Athena-XIFU at $7\ $keV  }  
\label{tab:xraymission}      
\centering                          
\begin{tabular}{c c c}       
\hline\hline                 
 Instrument & Resolving Power & Effective Area\\   
  & $E/\Delta E$ & [cm$^2$]\\
\hline                      
 Chandra HETG\tablefootmark{a} 	&    148      	& 18	\\
 XARM SXS\tablefootmark{b}  	&    1750      	& 225 	\\
 Athena X-IFU\tablefootmark{c}  &    2800      	& 1600 	\\
\hline                                   
\end{tabular}
\tablebib{(a) \citet{Canizares05};
(b) \citet{Mitsuda12}; (c) \citet{Barret16}
}
\end{table}

   \begin{figure}
   \centering
   \includegraphics[width=\hsize]{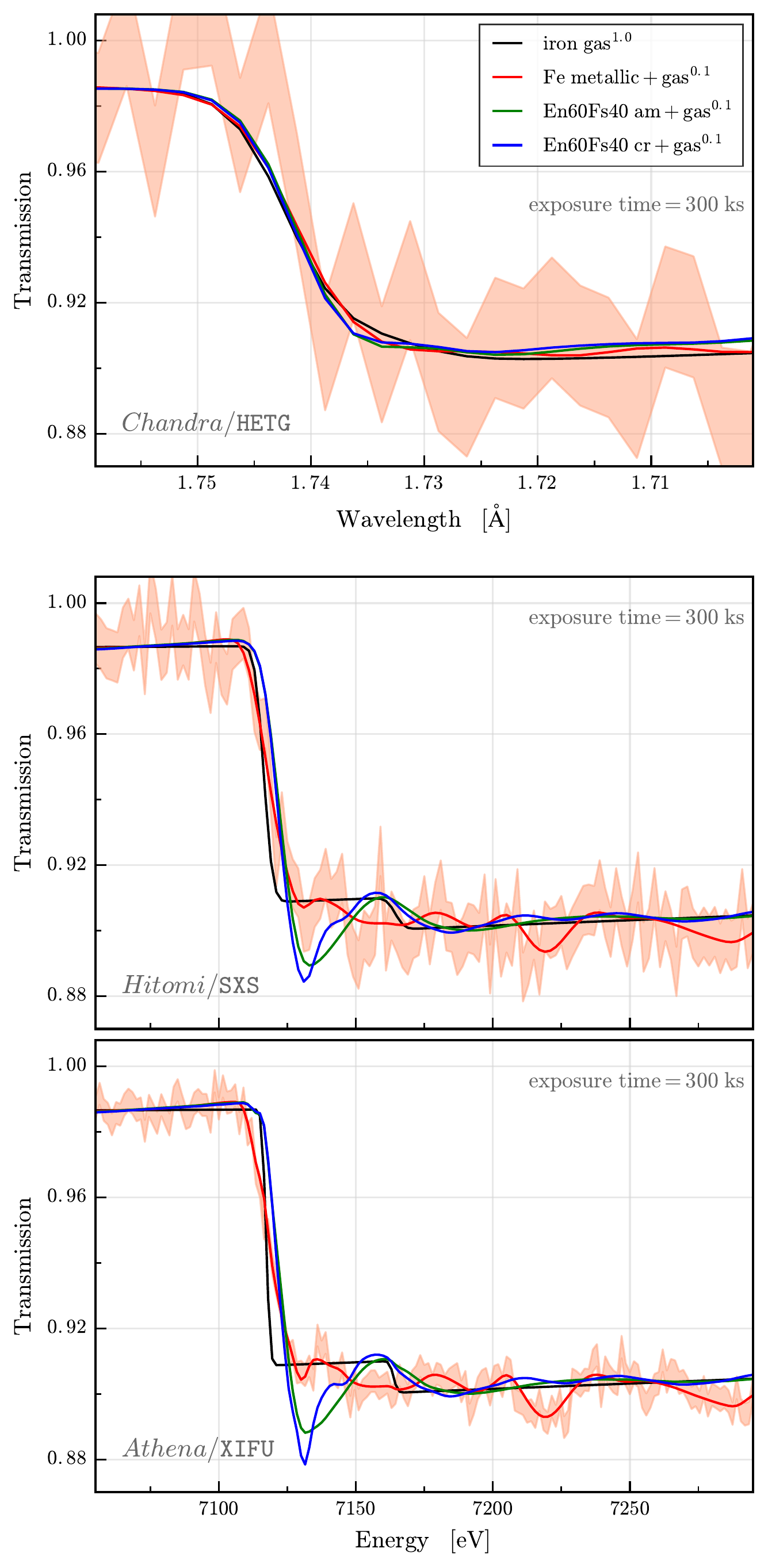}
      \caption{Simulated transmission for a mixture of gas ($10\%$) plus a dominant dust component: iron gas (black), metallic iron (red), and pyroxene En60Fs40 in both amorphous (green) and crystalline (blue) form. We set a solar abundance and we also simulate the transmission of pure iron gas (black) as a reference. The red shadow represents the noise for the iron metallic model. \emph{Top}: Simulation with Chandra-HETG. \emph{Middle}: Simulation with the XARM-SXS microcalorimeter \citep[see also][]{Paerels14}. \emph{Bottom}: Simulation with Athena-XIFU \citep[see also][]{Decourchelle13}.  
              }
         \label{fig:xarm_athena_chandra}
   \end{figure}

\subsection{Simulated sources}

X-rays are highly penetrating, thus it is possible to explore the dust in the denser regions of the Galaxy using low mass X-rays binaries as background sources. The presence of the absorption edges of O, Mg, Si, and Fe in the extinction profile depends on the column density ($N_{\text{H}}$) on the line of sight towards the source. It is difficult to detect dust in the Fe K-edge region. Extreme conditions are required to observe the iron absorption features: \emph{i)} a very high hydrogen column density $> 6 \times 10^{22}\ \text{cm}^{-2}$ and \emph{ii)} a bright background X-ray source with a $2-10\ $keV flux $> 1 \times 10^{-9}\ \text{erg}\ \text{cm}^{-2}\ \text{s}^{-1}$.\\
In order to test and compare the sensitivity of the present Chandra transmission grating and the future XARM and Athena microcalorimeters we simulate the response of these instruments under optimised conditions to study the Fe K-edge for a reasonable exposure time. We simulate an outbursting X-ray source near the Galactic centre \citep[here SAXJ1747.0-2853,][]{Zand04} with a $2-10\ $keV flux $3\times10^{-9}\ \text{erg}\ \text{cm}^{-2}\ \text{s}^{-1}$ and $N_\text{H}=8.5\times 10^{22}\ \text{cm}^{-2}$. In this simulation we considered a mixture of gas (15\%) and one dominant dust component in amorphous or crystalline form (either Mg$_{0.6}$Fe$_{0.4}$Si$_3$ or metallic Fe) with the standard MRN dust size distribution. The total Fe abundance has been set to proto-solar \citep{Lodders10}. \\ 
As shown in Figure \ref{fig:xarm_athena_chandra}, the resolving power of Chandra-HETG is not enough to detect any dust features in the Fe K region. It will be possible to distinguish the iron silicate present in the dust with XARM-SXS because the silicate peak at $\sim 7130\ $eV can be easily detected. However, only the resolving power of Athena-XIFU is able to detect the secondary features and then identify the different compound with enough accuracy.

\subsection{Chemistry, depletion, and abundances}
\begin{figure}
   \centering
   \includegraphics[width=\hsize]{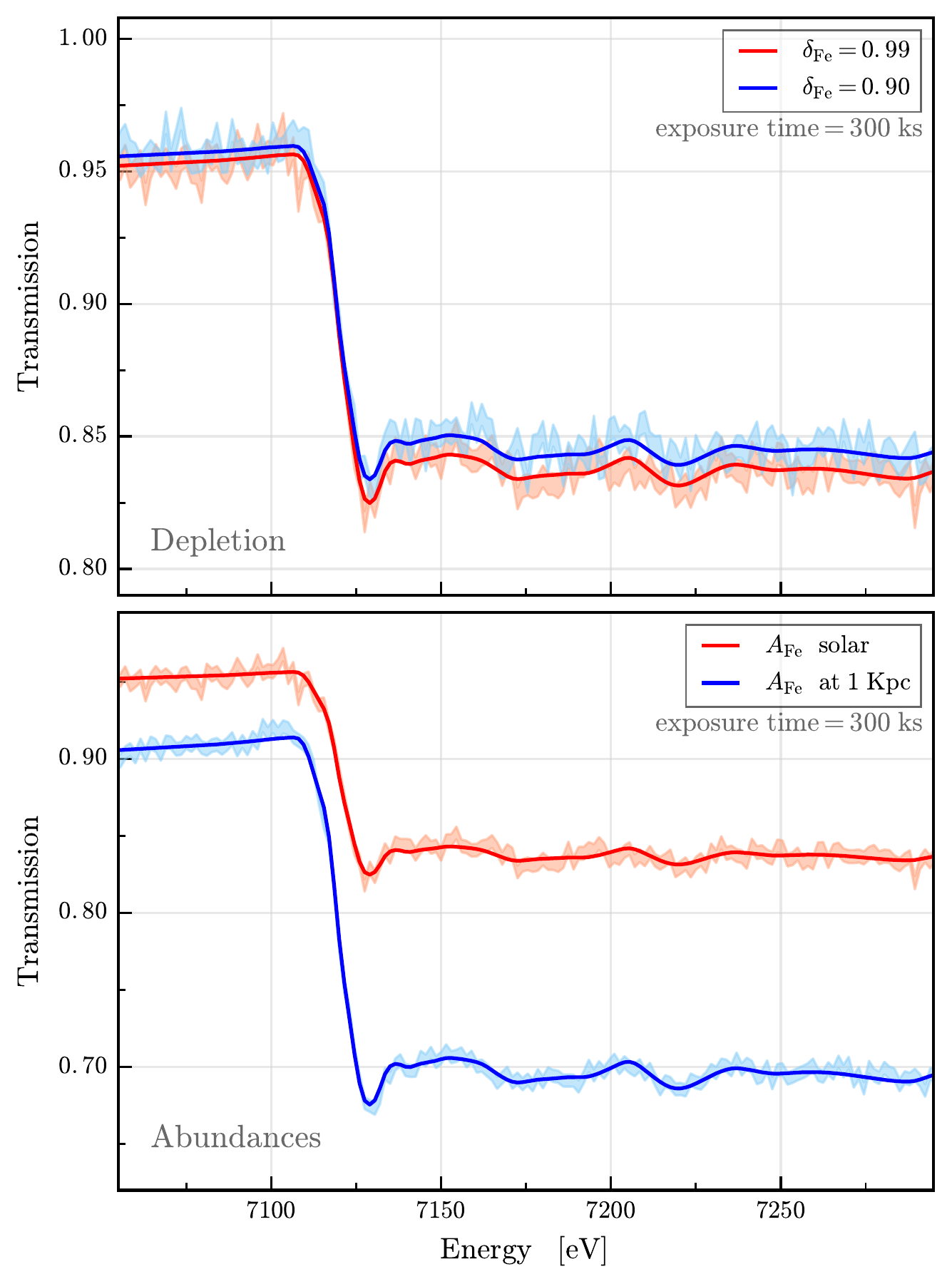}
      \caption{Depletion and abundances simulations with the Athena-XIFU instrument for the source and the hypothetical interstellar dust composition presented in the text. \emph{Top}: transmittance expected for two different iron depletion values: $\delta_{Fe} = 0.9$ in blue and $\delta_{Fe} = 0.99$ in red. We assume solar abundances. \emph{Bottom}: transmittance calculate taking into account solar abundance of iron (in red) and abundance expected in a region located at $1\ $kpc away from the Galactic centre (in blue). We fix the depletion at $\delta_{Fe} = 0.99$. 
              }
         \label{fig:depletion_abundance}
   \end{figure}

Here we test the sensitivity of the X-IFU to both depletion and abundances of iron. We consider a mixture of iron silicates, sulfides, and metallic iron (50\%, 40\%, and 10\%, respectively, of the total amount of iron included in dust). To develop the simulation we take into account the same source presented in the previous section.\\
Iron is heavily embedded in dust and its depletion factor ranges between 0.90 and 0.99 \citep{Savage96, Jenkins09}. In Figure \ref{fig:depletion_abundance}$a$ we present the transmission expected for the limit values of the depletion range. The two curves do not change significantly. The data errors overlap for almost the entire the edge region and this would complicate the calculation of the percentage of iron present in the solid and gaseous phase.\\
The iron abundance in the innermost Galactic disc regions is well above solar ($[\text{Fe}/ \text{H}]\sim 0.4$, \citeauthor{Pedicelli09} \citeyear{Pedicelli09}), while in the outer disc it is significantly more metal-poor ($[ \text{Fe}/ \text{H}] \sim -0.2/-0.5$, \citeauthor{Lemasle08} \citeyear{Lemasle08}). Figure \ref{fig:depletion_abundance}$b$ compares the transmission with solar abundance to the transmission with over-solar iron abundance for a source near the Galaxy centre. We fix the depletion factor to 0.99. To calculate the abundance of iron for a region located at $7.5\ $kpc away from the Earth and $1\ $kpc away from the Galactic centre we consider the iron gradient presented by \cite{Genovali14}. The simulations show a detectable difference especially in the depth of the Fe K-edge, which is larger for over-solar metallicities. 

\subsection{Grain size}
\begin{figure}
   \centering
   \includegraphics[width=\hsize]{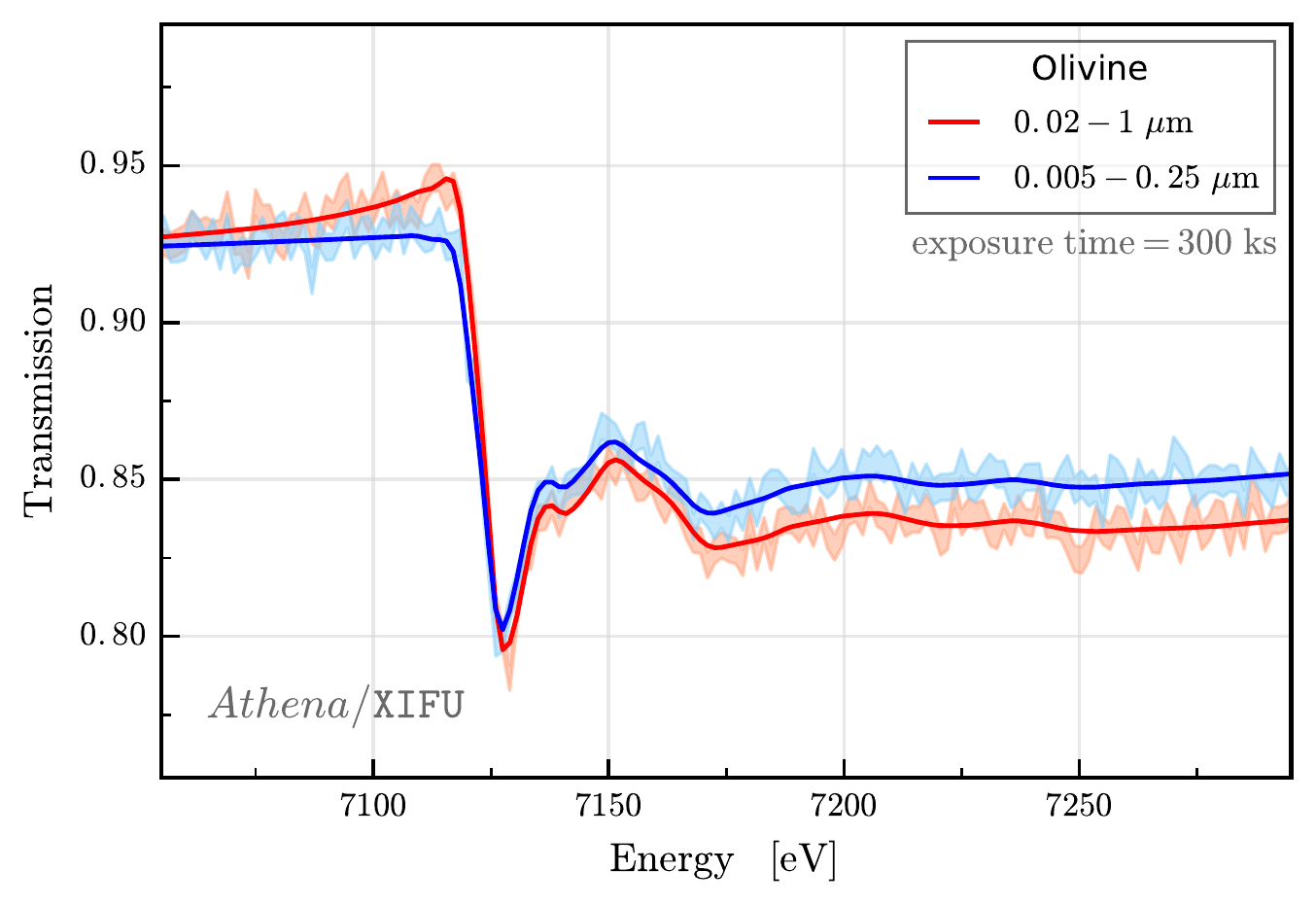}
      \caption{Simulated transmittance for two dust size ranges: ($0.005\le \rm{a} \le0.25)\ \mu$m (blue) and $(0.02\le \rm{a} \le1)\ \mu$m (red). 
              }
         \label{fig:large_grains_sim}
   \end{figure}

In Section \ref{sec:fek} we see how the pre-edge shape is sensitive to the dust size. Here we test the capability of Athena to investigate the expected grain size in the most dense region of the Galaxy. In Figure \ref{fig:large_grains_sim} we present the results of the simulation for the two size ranges. The population of grains with sizes ranging between $0.02$ and $1\ \mu$m shows a pronounced scattering peak in the pre-edge region. Consequently with Athena-XIFU we will be able to distinguish the peak and asses the presence a population of large grains along the line of sight.

\section{Conclusion}
\label{sec:con}
In this paper we present the laboratory measurements of silicates in amorphous and crystalline forms and iron sulfides that we carried out at the synchrotron facilities in transmission geometry. In our study we include the literature value of metallic iron. Using these new data, we calculated the scattering, absorption, and extinction profiles of these samples. We implemented the absolute extinction curves in the spectral model AMOL of SPEX. We then explored the potential of the Fe K-edge probing the interstellar dust properties. The results of this analysis are as follows:
\begin{enumerate}
\item The absorption cross section in the Fe K-edge region is very sensitive to the chemical composition of the dust grains. The post-edge is a real fingerprint of the grain chemistry.
\item The absorption features in the Fe K post-edge region are also sensitive to the lattice order of the material. Non-crystalline solids display a smoother curve than the crystalline samples. The transition from crystalline to amorphous phase does not affect the characteristics of the main features in the spectra, although it levels out the secondary peaks. In addition, the extinction cross section near the iron K-edge is insensitive to the grain porosity and geometry and to the grain size distribution model chosen. Thus, the iron K-edge spectra is not contaminated by other grain properties. 
\item The scattering cross section is significantly sensitive to the grain size range. For grains with sizes ranging between $0.005-0.25\ \mu$m (as in the MRN distribution) the scattering cross section is flat and does not affect the shape of the extinction cross section (see Figure \ref{fig:cross}). If we consider grains with sizes ranging between $0.02-1\ \mu$m, the scattering cross section becomes more intense; in particular, it shows a deep feature in the pre-edge region. The extinction cross section displays the scattering peak for large grains in the pre-edge region. By detecting this scattering feature, it will be possible to investigate directly the dust size range even in very dense regions of the Galaxy.     
\item Iron is highly depleted in the Galaxy and the level of the depletion does not change enough to vary the transmittance significantly. However, the edge is suitable for investigating the iron abundance gradient, which influences considerably the attenuation of the transmission and the depth of the Fe K-edge.
\end{enumerate}
The simulations with the XARM-SXS microcalorimeter and the Athena-XIFU spectrometer show the potential of using the Fe K-edge at $7.112\ $keV as a diagnostic tool. This edge will be a useful probe to investigate the iron abundance, the chemistry, and the grain size of the interstellar dust in the most dense environment of the Galaxy. These properties show distinctive features in the spectrum, making the modelling of different properties of dust easier. \\

\begin{acknowledgements}
DR and EC are supported by the Netherlands Organisation for Scientific Research (NWO) through \emph{The Innovational Research Incentives Scheme Vidi} grant 639.042.525. The NWO is acknowledged for making access to the DUBBLE beamlines possible. The Space Research Organization of the
Netherlands is supported financially by the NWO. We thank Hiroki Chihara and Simon Zeidler for their contributions to the preparation of the sample. Furthermore, we made use of \emph{KKcalc} code provided by Benjiamin Watts. We also thank A. Dekker for commenting on this manuscript. 
\end{acknowledgements}

   \bibliographystyle{aa} 
   \bibliography{aa}
   

\begin{appendix} 
\section{Extinction cross sections}
\label{app:samples}
\begin{figure}
   \centering
   \includegraphics[width=\hsize]{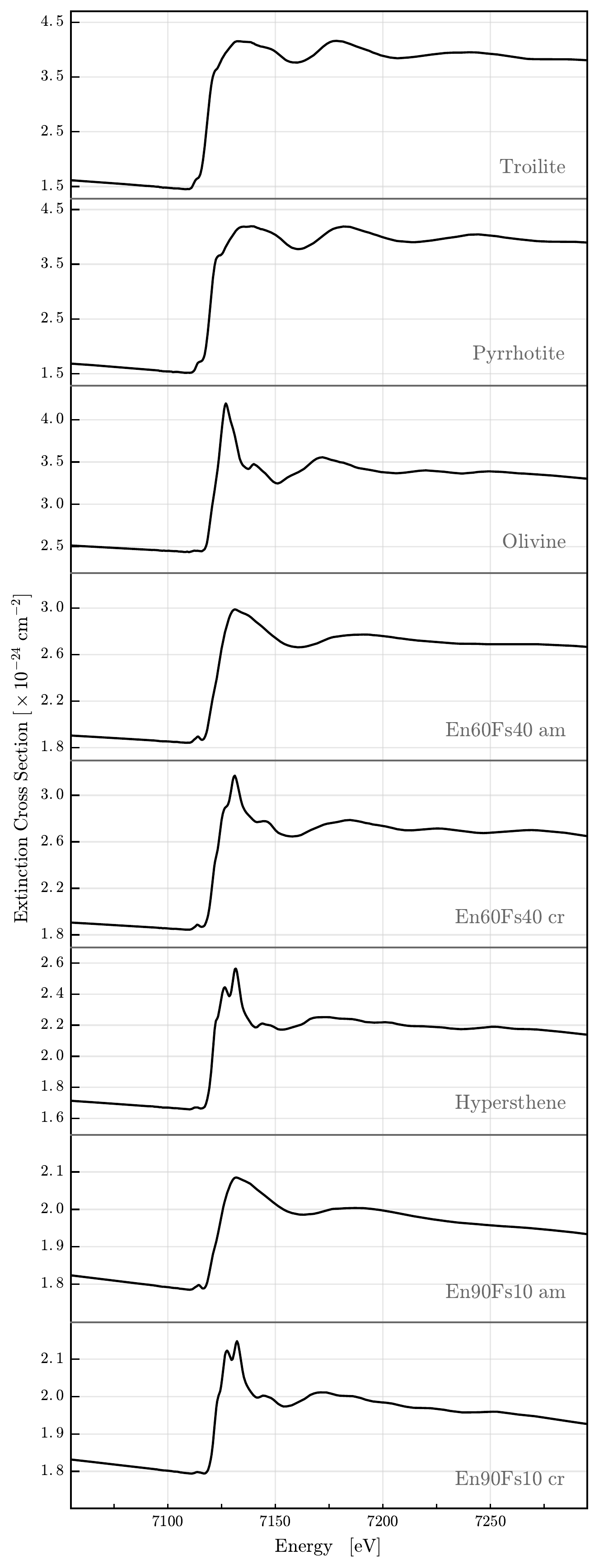}
      \caption{Iron K-edge extinction cross sections of the samples presented in this work.}
         \label{fig:sample}
   \end{figure}

In Figure \ref{fig:sample} we present the extinction cross section profiles around the iron K-edge for each sample presented in Table \ref{tab:sample}. We developed these data from the synchrotron measurements taken at the European Synchrotron Radiation Facility in Grenoble, France. These curves were implemented in the AMOL model of the spectral fitting code SPEX with a fixed energy resolution of $0.1\ \text{eV}$. The absorption, scattering, and extinction cross sections of the compounds (with an energy range between 6.7 and 8.0~keV) are available in ASCII format at the following link: \url{www.sron.nl/~elisa/VIDI/}.

\end{appendix}

\end{document}